\documentclass[twocolumn,showpacs,preprintnumbers,prb]{revtex4}

\usepackage{psfig}
\usepackage{dcolumn}
\usepackage{bm}

\begin{document}

\title{Polaron formation in the optimally doped ferromagnetic manganites La$%
_{0.7}$Sr$_{0.3}$MnO$_{3}$ and La$_{0.7}$Ba$_{0.3}$MnO$_{3}$}
\author{Y. Chen}
\affiliation{NIST Center for Neutron Research, National Institute of Standards and
Technology, Gaithersburg, MD 20899}
\affiliation{Dept.\ of Materials Science and Engineering, University of Maryland, College
Park, MD 20742}
\author{B. G. Ueland}
\affiliation{NIST Center for Neutron Research, National Institute of Standards and
Technology, Gaithersburg, MD 20899}
\author{J. W. Lynn}
\affiliation{NIST Center for Neutron Research, National Institute of Standards and
Technology, Gaithersburg, MD 20899}
\author{G. L. Bychkov}
\affiliation{Institute of Solid State and Semiconductor Physics, Academy of Science,
Minsk 220072, Belarus}
\author{S. N. Barilo}
\affiliation{Institute of Solid State and Semiconductor Physics, Academy of Science,
Minsk 220072, Belarus}
\author{Y. M. Mukovskii}
\affiliation{Moscow State Steel and Alloys Institute, Moscow 119049, Russia}
\date{\today}

\begin{abstract}
The nature of the polarons in the optimally doped colossal magnetoresistive
(CMR) materials La$_{0.7}$Ba$_{0.3}$MnO$_{3}$ (LBMO) and La$_{0.7}$Sr$_{0.3}$%
MnO$_{3}$ (LSMO) is studied by elastic and inelastic neutron scattering. In
both materials, dynamic nanoscale polaron correlations develop abruptly in
the ferromagnetic state. However, the polarons are not able to lock-in to
the lattice and order, in contrast to the behavior of La$_{0.7}$Ca$_{0.3}$MnO%
$_{3}$. Therefore ferromagnetic order in LBMO and LSMO survives their
formation, explaining the conventional second order nature of the
ferromagnetic--paramagnetic transition. \ Nevertheless, the results
demonstrate that the fundamental mechanism of polaron formation is a
universal feature of these ferromagnetic perovskite manganites.
\end{abstract}


\maketitle

\section{INTRODUCTION}
The discovery of colossal magnetoresistance (CMR) in the hole-doped
perovskite manganites of the form La$_{1-x}$A$_{x}$MnO$_{3}$ (A=Ca, Ba and
Sr) has generated much interest \cite{DagottaReview, TokuraReview}. The
parent compound LaMnO$_{3}$ is an antiferromagnetic insulator, with a Mn-O
sublattice consisting of a network of corner-sharing Mn$^{3+}$O$_{6}$
octahedra. The Mn$^{3+}$ ion has three $t_{2g}$ electrons that form its
S=3/2 core-spin and one additional electron in its higher energy,
doubly-degenerate $e_{g}$ orbital. The Jahn-Teller distortion of the MnO$%
_{6} $ octahedron removes the degeneracy of the $e_{g}$ levels and the
ground state consists of an ordered array of distorted octahedra with
antiferromagnetic order of the Mn spins. With substitution of trivalent La$%
^{3+}$ by divalent alkaline-earth metal ions such as Ca$^{2+}$, Ba$^{2+}$,
or Sr$^{2+}$, holes are introduced into the material, producing Mn$^{4+}$
ions where the $e_{g}$ orbital is unoccupied. With sufficient hole doping
(typically $0.2<x<0.5$) the ground state becomes a magnetically isotropic
ferromagnetic metal. In this regime the basic relationship between
conductivity and ferromagnetism has been understood by the double-exchange
(DE) mechanism \cite{Zener}, in which itinerant $e_{g}$ electrons associated
with the Mn$^{3+}$ hop between the localized Mn$^{4+}$ ions, providing both
ferromagnetic exchange (due to the strong Hund's rule coupling between the $%
t_{2g}$ and $e_{g}$ electrons) and electrical conduction.  The DE interaction competes with a strong electron-phonon coupling via the Jahn-Teller active 
Mn$^{3+}$ ions.  The hopping of an electron, with its associated lattice distortion, forms a polaron.  
  Here we focus on the regime near optimal doping for ferromagnetism, where CMR is observed.  At optimal doping, this competition produces a transition from the ferromagnetic-metallic ground state to an intrinsically inhomogeneous paramagnetic-insulating state at elevated temperatures, which is often described as polaronic \cite{Millis1995, Millis1996}.

 For the prototype La$_{0.7}$Ca$%
_{0.3}$MnO$_{3}$ (LCMO) system the ferromagnetic state has been found to be
truncated by the formation of static nanoscale lattice polarons,\cite{Adams2000,
Dai2000,Lynn2001, Nelson2001, Kiryukhin2004} driving the ferromagnetic-paramagnetic
transition first-order\cite{Lynn1996, Adams2000, Adams2004, Kim2002, Rivadulla2006,Lynn2007,Demko2008}.  These polarons\cite{Downward2005, Souza2008} make the system especially sensitive to applied magnetic fields, amplifying the
magnetoresistivity near $T_{C}$.  In LCMO, polaron correlations develop within a narrow temperature regime as $T_{C}$ is approached from low temperatures, with a nanoscale correlation length that is only weakly temperature dependent.  Polarons take the form of CE-type correlations with an ordering wave vector of $\approx$ (1/4,1/4,0) (with respect to the cubic perovskite cell).   The static nature of these short range polaron correlations indicates the presence of a glass-like state \cite{Adams2000,Dai2000,Adams2004}.  In addition, dynamic correlations exist with a comparable correlation length and with an energy distribution that is quasielastic.  The elastic polaron scattering disappears at higher temperature, above which the correlations are purely dynamic \cite{Lynn2007}.  While these CE correlations characterize the insulating state above $T_{C}$, the onset of ferromagnetism leads to their melting.     

In the closely related La$_{0.7}$Ba$_{0.3}$MnO$_{3}$ (LBMO)\cite{Barilo2000, Jiang2008} and La$_{0.7}$Sr$_{0.3}$MnO$_{3} $ (LSMO)\cite{Ghosh1998, Doloc1998} ferromagnets, on the other hand,
the ferromagnetic transition is a conventional second-order one, and no
polaron order is observed. \ In the present work we have found that
nanoscale polarons do indeed form in optimally doped LSMO and LBMO, well
below $T_{C}$, in the ferromagnetic state. \ However, these polarons are
purely dynamic in nature, and never lock-in to the lattice to form an
ordered state. \ The ferromagnetic state therefore survives their formation,
and a conventional second-order transition is realized at higher
temperatures, well above the temperature where the polarons form. \ The
present results demonstrate that intrinsic inhomogeneities on a nanometer
length scale are a universal feature of these ferromagnetic manganites,
while the ferromagnetic state is only terminated if the polarons condense to
form a (static) glass phase.\cite{Argyriou2002,Lynn2007}

\section {EXPERIMENTAL PROCEDURES}
Elastic and inelastic neutron scattering measurements were carried out to
study the polaron correlations in single crystals of La$_{0.7}$Sr$_{0.3}$MnO$%
_{3}$ and La$_{0.7}$Ba$_{0.3}$MnO$_{3}$ weighting approximately 3 and 1
grams, respectively. These crystals have second-order ferromagnetic
transitions as evidenced by the temperature dependence of the magnetic Bragg
peaks as well as the small wave vector spin wave dispersion data, and are
the same crystals used in earlier studies, with $T_{C}$'s of 351 K\cite%
{Doloc1998} and 336 K,\cite{Barilo2000} respectively. The neutron data were
obtained on the BT-7 and BT-9 thermal triple axis spectrometers at the NIST
Center for Neutron Research.  We used pyrolytic graphite (PG) (002) monochromators and PG (002) analyzers.  PG filters of 5 cm thickness were used to eliminate higher-order contamination.  For BT-7 measurements, horizontal collimators with divergences of 100$^\prime$, 50$^\prime$, 50$^\prime$ and 100$^\prime$ full width at half maximum (FWHM) were used before and after the monochromator, after the sample, and before the detector, respectively.  For BT-9 measurements, collimations were 40$^\prime$-47$^\prime$-40$^\prime$-80$^\prime$.  Samples were
sealed in an aluminum can containing helium exchange gas and mounted in a
closed cycle refrigerator operating over a temperature range from 10 K to
400 K.  The crystal structure of LSMO and LBMO is orthorhombic, but only
slightly distorted from the cubic lattice, and since the crystals are
twinned for simplicity we label wave vectors in the reciprocal lattice units
appropriate for pseudocubic unit cells with lattice parameters of a = 3.8660 
\AA ~and a = 3.9232 \AA ~(at 360 K) for LSMO and LBMO, respectively. \ In
this notation the polaron correlations are observed around (1/4,1/4,0) and
equivalent positions (CE-type lattice correlations), in reciprocal lattice
units (rlu) defined by $a\ast =2\pi /a$. 

\section{EXPERIMENTAL RESULTS}
 
\begin{figure}[tbp]
\centerline{\psfig{file=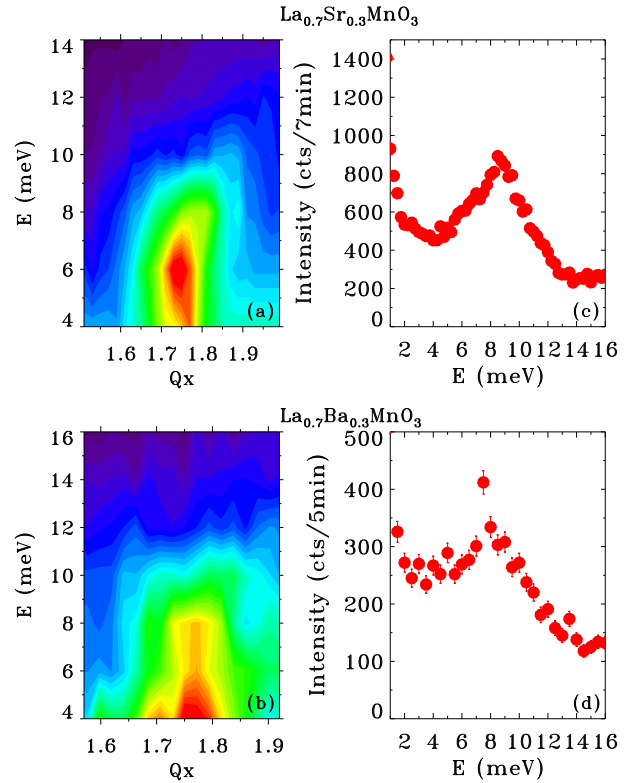,width=1.\columnwidth,angle=0,clip=}}
\caption{(color online) ($\mathbf{Q}$,E) map of inelastic neutron scattering
intensity around $\mathbf{Q}$ = (1.5+$q$,2+$q$,0) (cubic notation) at $T$ =360 K for (a) LSMO
and (b) LBMO. Dynamic polaron scattering is observed around the q=
(1/4,/1/4,0) reduced wave vector position. (c) and (d) show an energy scan
at $\mathbf{Q}$ = (1.75,2.25,0) for LSMO and LBMO, respectively, at $T$ = 360 K,
revealing that the scattering has a broad maximum at finite energy, in
addition to a quasielastic component.  The energy resolution at the elastic position is 1.5 meV.  \ Error bars where shown\ are
statistical in nature and represent one standard deviation.}
\label{dynamic}
\end{figure}

Figure \ref{dynamic} shows a color map of the neutron intensity above $T_{C}$
for each system as a function of ($\mathbf{Q}$,E). We see that the dynamic
correlations are observed and are peaked around the $\mathbf{Q}$ =
(1.75,2.25,0) position, as expected for these CE-type lattice correlations.
\ The intensity for LSMO has a broad maximum in energy, which can be seen
more clearly in the energy scan shown in Fig. \ref{dynamic}(c). For LBMO
this maximum occurs at a somewhat lower energy (Fig. \ref{dynamic}(d)). \
These data were taken with the same instrumental conditions, and recalling
that the LSMO sample is 3$\times $ larger than the LBMO sample, we see that
the strength of the polaron scattering is comparable in the two systems. The
scattering extends to low energies, and appears to consist of two
components, a quasielastic component (a Lorentzian centered at E=0) and a
component that has a broad maximum at finite energy. \ At higher energies
the scattering decreases fairly rapidly. \ The overall behavior is quite
similar to the dynamics observed in LCMO\cite{Lynn2007}, where again there
is a quasielastic component to the scattering along with a component that
peaks at finite energy. \ In that case it was difficult to extract a
quasielastic linewidth vs. temperature, in contrast to what was done for the
lower $T_{C}$ bilayer system.\cite{Argyriou2002} This is because of the
difficulty in establishing `background' at these elevated temperatures, as
well as the difficulty in separating the (apparent) two components of the
scattering. \ In the present case the $T_{C}$'s are even higher than for
LCMO, and hence it is even more difficult to extract detailed quasielastic
linewidths as a function of $T$. The general behavior, on the other hand, is
clear.

\begin{figure}[tbp]
\centerline{\psfig{file=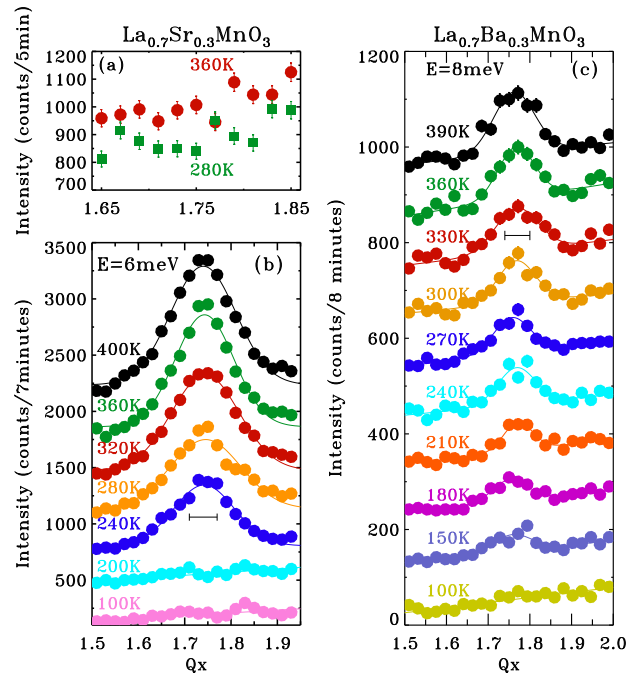,width=1.\columnwidth,angle=0,clip=}}
\caption{(color online) (a) Elastic scan through the polaron position for
LSMO above and below $T_{C}$. \ No static polaron peaks were observed at any
of the (1/4,1/4,0) positions for either LSMO or LBMO. (b) Temperature
dependence of the inelastic scattering for LSMO at an energy transfer of E =
6 meV for $\mathbf{Q}$ = (1.5+$q$,2+$q$,0). The scattering at successive temperatures has
been displaced vertically by 280 counts for clarity. A peak in the
scattering abruptly appears around $T$ = 240 K, and persists to higher
temperatures.  The horizontal bar indicates the q-resolution (FWHM). (c) Temperature dependence of the inelastic scattering in LBMO
at an energy transfer of E = 6 meV for $\mathbf{Q}$ = (1.5+$q$,2-$q$,0). The scattering at
successive temperatures has been displaced vertically by 100 counts for
clarity. A peak in the scattering abruptly appears around $T$ = 150 K, and
persists up to high temperature.  The horizontal bar indicates the q-resolution (FWHM).}
\label{tdep}
\end{figure}

\begin{figure}[tbp]
\centerline{\psfig{file=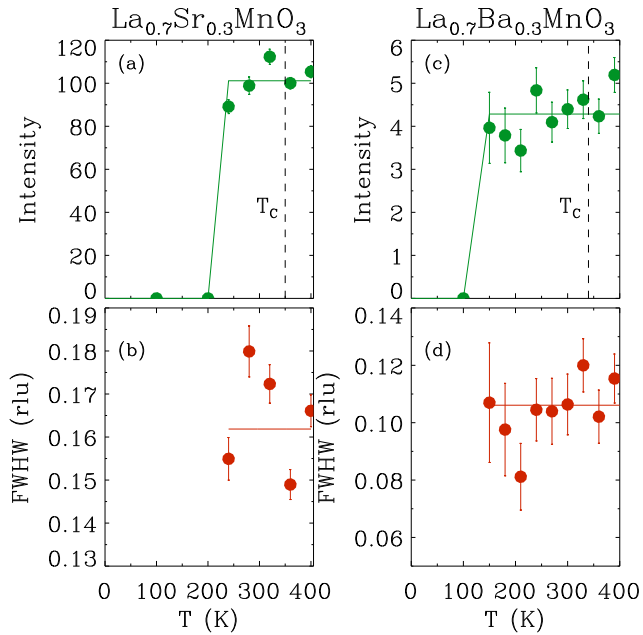,width=1.\columnwidth,angle=0,clip=}}
\caption{(color online) [(a) and (c)] Temperature dependence of the inelastic
scattering intensity shown in Fig. \protect\ref{tdep} (a) for LSMO and (c)
LBMO. The integrated intensity of the scattering abruptly appears at a
temperature well below $T_{C}$, and then is constant at higher temperature
after the Bose-Einstein thermal population factor has been divided out.
(b,d) Temperature dependence of the intrinsic width in Fig. \protect\ref{tdep}
for (b) LSMO and (d) LBMO. The width is found to be temperature independent.
\ Neither the width nor the intensity are sensitive to the ferromagnetic
ordering.}
\label{fitpara}
\end{figure}

To investigate the temperature dependence of the dynamic polaron
correlations, $\mathbf{Q}$ scans around the (1/4,1/4,0) position were
carried out for a series of energy transfers over a wide temperature range.
The dynamic polaron scattering is very broad in energy, and broader than the instrumental resolution in $\mathbf{Q}$ as well.  Hence, it is generally better to use $\mathbf{Q}$-scans to measure the scattering since the scattering is confined to a relatively small range and the resolution does not vary over the scan.  Both the $\mathbf{Q}$-width and energy width of the measured peaks are not resolution limited, and $\mathbf{Q}$-scans can be performed in a wide energy range to observe the dynamic polaron correlations.  Fig. \ref{tdep} shows data as a function of $\mathbf{Q}$ = (1.5-$q$,1.5+$q$,0) for
LSMO. \ To search for static CE-type polaron correlations, extensive elastic
scans (E = 0) were performed above and below $T_{C}$. No evidence of ordered
polarons was observed at any temperature, as indicated in the data shown in
Fig. \ref{tdep}(a). \ This contrasts with the behavior of LCMO, where
short-range static CE-type polaron correlations abruptly develop at $T_{C}$,%
\cite{Adams2000, Dai2000} and truncate the ferromagnetic state.\cite%
{Adams2004, Lynn2007}

The inelastic scattering, on the other hand, conveys a completely different
story. At $T$ = 100 K and 200 K there is no evidence in LSMO of any peak in
the scattering, with featureless scattering that is weakly temperature
dependent and likely originates from multiphonon and multimagnon
(background) scattering. \ At 240 K, however, a distinct peak centered at $%
\mathbf{Q}$ $\approx $ (1.75,2.25,0) abruptly appears. Note that this
temperature is far below $T_{C}$. The wave-vector width of this peak is
approximately temperature and energy independent and corresponds to a correlation
length of $\approx $ 1 nm (10 \AA ). Similar behavior is observed for LBMO,
where Fig. \ref{tdep}(c) shows $\mathbf{Q}$ scans measured at an energy
transfer of E = 8 meV between $T$ = 100 K and 390 K. A peak in the
scattering is observed at and above 150 K, while no peak is observed at
lower $T$. \ Hence in both systems we observe the abrupt development of
short-range dynamic polaron correlations well below $T_{C}$, which evolve
through $T_{C}$ without any apparent indication that long range
ferromagnetic order has been lost. These dynamic correlations persist up to
high temperatures, likely all the way up to the orthorhombic-rhombohedral
transition.\cite{Souza2008, Kiryukhin2004}

The data in Fig. \ref{tdep} have been fit to a linear background plus a
Gaussian peak. \ Once the peak appears, its observed intensity increases
with increasing $T$ and follows the thermal occupancy factor for a boson
excitation. \ The results for the integrated intensities, after correcting
for the Bose thermal factor, are shown in Fig. \ref{fitpara}. \ Fig. \ref%
{fitpara}(a, c) shows that the inelastic scattering abruptly appears at a
well defined temperature, and then is constant in strength. \ Therefore this
overall scattering behaves as a phonon excitation. \ The width of the
scattering (Fig. \ref{fitpara}(b,d)), once the intensity appears, is also
independent of temperature. \ Hence the correlation length in real space
remains at $\approx 1nm$ throughout the entire temperature range where this
scattering is observed. \ 
\section{DISCUSSION and CONCLUSIONS}

The development of static CE-type correlations has been shown to have
important implications for the ferromagnetic to paramagnetic transition in
the CMR manganites. When static polarons form in LCMO, electrons become
trapped, driving the system into an insulating state (hopping conductivity).
The transition into the insulating state is accompanied by a truncation of
the ferromagnetic state in a discontinuous fashion \cite{Adams2004}. Above $%
T_{C}$, the nature of the order is a polaron glass, in that the correlations
are short range, and static.\cite{Argyriou2002, Lynn2007} The short range
correlations make the system especially sensitive to modest magnetic fields
that can then drive the paramagnetic-insulating--ferromagnetic-metallic
transition, with the concomitant large change in electrical conductivity \cite{Downward2005, Souza2008}.  At sufficiently high $T$ the glass order disappears while dynamic polaron correlations persist as a polaron liquid.  In LCMO, we observed a sharp phonon at 8 meV in the ferromagnetic metallic state below $T_{C}$ \cite{Lynn2007}.  Above $T_{C}$, the phonon is still present, 
and interacts with the dynamic polarons.   However, no such phonon is observed in this energy range in LSMO and LBMO.  Hence, the observed scattering is purely from the polaron correlations.

In the course of the initial studies on LCMO,\cite{Adams2000} a search for
these polarons peaks was made in LBMO and LSMO, but no static polaron
correlations were found. Indeed for the latter materials the ferromagnetic
transition is a conventional second order one, with properties like the
magnetization and spin wave stiffness quantitatively exhibiting scaling
behavior,\cite{Ghosh1998, Jiang2008, Doloc1998, Barilo2000} while the rather
large changes in the resistivity can be understood on the basis of
conventional electron scattering by magnetic fluctuations.\cite{FisherLanger}
The behavior of the resistivity is also different than for LCMO, which peaks
at $T_{C}$ when the polarons develop. For LSMO $\rho (T)$ increases with $T$
(i.e. is metallic-like) both below and above $T_{C}$,\cite{Urushibara} while
for LBMO $\rho (T)$ peaks, but only well above $T_{C}$.\cite{Barilo2000}
The polarons in LBMO and LSMO abruptly develop, but well below $T_{C}$, and
there is no obvious effect on the resistivity. \ The sizes of the polarons
in all three systems are very similar as shown in Table 1, where the size
has been taken from the width of the inelastic scattering. \ We note that
the strength of the scattering, which is related to the number of polarons
(since the size is temperature independent), is also comparable in the three
systems. \ Hence it is not clear why only the polarons in LCMO lock-in to
the lattice and truncate the ferromagnetic state, even though the polarons
in LBMO and LSMO form at lower temperature (particularly in comparison to $%
T_{C}$) but never lock-in.\ It is also not clear what is controlling the
temperature of formation of the polarons, either static or dynamic, in these
ferromagnetic metallic systems. \ We hope that the present studies will
stimulate theoretical studies to address these issues.

\begin{table}[tbp]

\begin{tabular}{cccc}
\hline
\hline
 & Ca & Sr & Ba \\ \hline
$a(\mathring{A})$ & 3.868 & 3.866 & 3.923 \\ 
$T_{C}(K)$ & 257 & 351 & 336 \\ 
$\xi (\mathring{A})$ & 11.1(3) & 7.7(3) & 11.3(3) \\ \hline
\hline
\end{tabular}
\caption{Comparison of the lattice parameters, ferromagnetic transition
temperatures, and size of the polarons for LCMO, LBMO, and LSMO.}
\label{table}
\end{table}
\section*{ACKNOWLEDGMENTS}
We thank Elbio Dagotto, Dimitri Argyriou,\ and Adriana Moreo for stimulating
discussions. \ Work in Belarus was supported in part by BRFFI, Grant No.
F-08R-177.

\end{document}